# A ROLES-BASED COMPETENCY FRAMEWORK FOR INTEGRATING ARTIFICIAL INTELLIGENCE (AI) IN ENGINEERING COURSES

**J Schleiss**[1]
Otto von Guericke University Magdeburg
Magdeburg, Germany
https://orcid.org/0009-0006-3967-0492

**A Johri**
George Mason University
Fairfax, USA
https://orcid.org/0000-0001-9018-7574

## ABSTRACT

In this practice paper, we propose a framework for integrating AI into disciplinary engineering courses and curricula. The use of AI within engineering is an emerging but growing area and the knowledge, skills, and abilities (KSAs) associated with it are novel and dynamic. This makes it challenging for faculty who are looking to incorporate AI within their courses to create a mental map of how to tackle this challenge. In this paper, we advance a role-based conception of competencies to assist disciplinary faculty with identifying and implementing AI competencies within engineering curricula. We draw on prior work related to AI literacy and competencies and on emerging research on the use of AI in engineering. To illustrate the use of the framework, we provide two exemplary cases. We discuss the challenges in implementing the framework and emphasize the need for an embedded approach where AI concerns are integrated across multiple courses throughout the degree program, especially for teaching responsible and ethical AI development and use.

---

[1] *Corresponding Author*
*J Schleiss*
johannes.schleiss@ovgu.de

# 1 INTRODUCTION

Computing and computing education have become an integral component of the engineering profession and engineering education. Topics such as programming, data analysis, and computational sciences are now commonly taught in the engineering curricula (Malmi and Johri 2023; Raj et al. 2019). In recent times, the field of artificial intelligence (AI), encompassing domains like data mining, machine learning, natural language processing, large language models, and transformers, has also begun to impact engineering (Broo, Kaynak, and Sait 2022; Johri 2020; Mustapha, Yap, and Abakr 2024). Consequently, in addition to computing literacy or computational thinking skills, it has become necessary to integrate these topics within engineering curricula and pedagogy (Magana, Falk, and Reese Jr 2013).

Given the expansiveness and ever-evolving nature of AI as a field with diverse applications, comprehending the specifics of AI education, including what and how they should be taught, is a challenging task for educators. Recently, there have been several efforts towards defining AI literacy as a basic set of competencies integral to for acting in an AI-driven professional environment (Long and Magerko 2020; Laupichler et al. 2022). Several studies have proposed generic competencies in AI (Laupichler et al. 2022; Ng et al. 2021) that cut across disciplines but increasingly there is an emphasis towards a more discipline-based approach. For instance, Stolpe and Hallström (2024) proposed AI literacy for technology education, Knoth et al. (2024) outlined an assessment matrix spanning from generic to domain-specific AI literacy and Schleiss et al. (2022) proposed an interdisciplinary competence profile for AI in engineering, that highlights the intersection of disciplinary and AI knowledge and skills.

While several teaching approaches and implementations have been proposed and tested (Isaac Flores-Alonso et al. 2023; Singelmann and Covarrubias 2023; Ng et al. 2023), the adoption of AI literacy in engineering education has not received much focus. Consequently, educators are challenged with identifying relevant competence constructs in the context of their disciplinary domain. Since AI in engineering is largely application-oriented, it is important for educators to understand the disciplinary and interdisciplinary contexts of these applications. At the same time, educators in engineering education do not always have a background in AI and their domain, making it difficult to identify the relevant competencies and building a barrier to integrating AI education into the disciplinary teaching offers.

To address the gap of clarity in relevant AI competencies, we propose a role-based approach that integrates AI literacy and competency to assist disciplinary faculty with identifying and implementing AI competencies within engineering curricula. In particular, we focus on the question of how to identify and select relevant AI competencies for discipline-specific AI education in engineering education. The paper builds on prior work related to AI literacy and competencies within the literature and on emerging research on the use of AI in engineering. To situate prior AI literacy frameworks within an engineering context, we take a roles-based approach. Our framework focuses on the use of AI by a specific actor within the engineering context and how this leads to the need for integration of a distinct AI competency within the curriculum. We inductively analyze case studies of the use of AI in engineering to understand emerging patterns and translate findings from literature into a practical and usable framework for educators.

Overall, this study aims to contribute to integrating AI education in engineering education across disciplines (Laupichler et al. 2022; Schleiss et al. 2023; Patel et al. 2021). In the following, we first introduce related work on AI literacy definitions and competencies and the roles-based approach that provide the base to our framework. Next, we describe the cases and inductively develop the **r**oles-based **c**ompetency for **AI** in **e**ngineering or RCAIE framework. Last, we discuss our findings and provide conclusions.

## 2 RELATED WORK

### 2.1 AI Literacy Definitions and Competencies

With the field of AI literacy being a novel and growing research field, there exist many definitions of AI literacy aiming at different target populations, purposes or from different disciplinary perspectives (Long and Magerko 2020; Laupichler et al. 2022; Ng et al. 2021; Stolpe and Hallström 2024; Knoth et al. 2024). Most prominent is the definition of Long and Magerko (2020) who define AI literacy as "a set of competencies that enables individuals to critically evaluate AI technologies, communicate and collaborate effectively with AI, and use AI as a tool online, at home, and in the workplace" (p.2). To bridge the gap between definitions of AI literacy to educating these competencies, several conceptual frameworks exist (Almatrafi, Johri, and Lee 2024). In their literature review, Almatrafi, Johri, and Lee (2024) identified six core constructs that form AI literacy: (1) *Recognize (Be Aware)* different types of AI applications as a basis to enable informed interactions, (2) *Know and Understand* fundamental concepts and techniques of AI, (3) *Use and Apply AI* applications and tools to solve tasks or achieve an objective, (4) *Evaluate* as the "ability to analyze and interpret the outcomes of AI applications critically", (5) *Create* as the ability to design and code AI applications, (6) *Navigate ethically* refers to understand ethical and social implications and become a responsible user of AI. These categories form an understanding of the core constructs of AI literacy and can be used as an analysis lens as well as to identify disciplinary-specific AI literacy constructs. Given the extensive systematic review on which this list of constructs rests, we utilize them as a core part of our framework but extend them towards domain-specific competencies in the context of roles and use-cases in engineering education.

### 2.2 Roles-based Approach to Competencies

Roles-based approach, although new to engineering, has widespread acceptance within another professional field – medical education. The approach has been applied both to roles in the profession that students need to be prepared for and for better understanding roles played by faculty as educators. For example, Harden and Crosby (2000) argued that given the changing and complex role of teaching in medical education, it was important to categorize these roles so that teachers or educators could identify what kind of expertise they needed in the medical field and in their roles as teachers. Whitehead et al. (2014), further argue that although more work needs to be done, the use of roles to describe competencies and build competency-based frameworks is now the norm in medical education reflecting "a conceptual shift in understandings of medical education from a process representing one of knowledge acquisition and time-based clinical rotations, to one in which competence can be considered as the adequate performance of a set of professional roles" (pg. 786). Similarly, Cenkner et al. (2017) identify six roles for an educational technology expert and for each role describe the corresponding competencies and give examples of activities that someone playing that role engages in.

In summary, a roles-based approach makes it easier for those who are new to a topic or area to approach it and have a guide or mental model of what it looks like including the responsibilities and objectives. The encapsulation of the knowledge as a role makes it task based and thereby easier for a newcomer to grasp given the alignment with practice. For this paper, we have created roles as group-level constructs meaning that roles define or identify groups of professionals who are "guided in their domain of practice by an established set of heuristics for thought and action (McClarey 2004; pg. 4)."

Professionals performing a certain role require a specific set of knowledge or expertise. The knowledge needed to perform a given role is known as 'role knowledge' (McClarey 2004; pg. 12). We adopt this approach towards examining roles that professionals in the workplace can play in relation to AI and what this means for what students need to be taught to be prepared for the workforce. In this initial mapping, we develop a broad conceptualization that can be expanded and defined more precisely through future work.

Within the literature on AI literacy, a roles-based approach has found resonance. Faruqe, Watkins, and Medsker (2021) distinguish between four groups of users with different depth of contact and competency requirements: (1) consumers, the general public and policymakers, (2) co-workers and users of AI products, (3) Collaborators and AI Implementers, and (4) creators of AI. Similarly, Schüller et al. (2023) distinguish between (1) informed prosumer, (2) skilled user, and (3) expert creator and developer. These roles correspond with different skill profiles within the industry such as AI Management, ML Architect, Data Engineer, Data Governor, Product Owner, Data Analyst, Data Architect, Data Steward, Data Scientist and ML Engineer (Machado and Mynter 2024). Given that this is an emergent area and there has been no rigorous analysis of AI-related roles for engineers, we synthesized three broad roles that we believe can provide guidance for knowledge assessment and skills development: a) consumer; b) user; and c) creator and developer. These roles in conjunction with the AI literacy constructs stated above provide us with a framework to describe the knowledge needed for AI integration within engineering.

## 3   RCAIE Framework and Case Studies

Overall, the **R**oles-based **C**ompetency in **AI** for **E**ngineering (RCAIE) framework builds on three pillars of use cases, roles and competencies (see Figure 1) and acts as foundation to identify suitable learning experience.

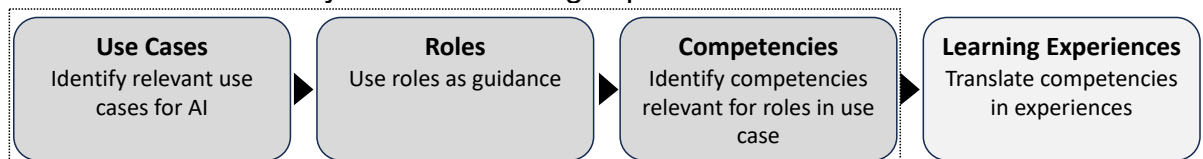

*Figure 1: Process to identify competencies with the RCAIE framework*

Given the wide application area for AI in engineering, we selected two exemplary case studies that help us explicate and describe the RCAIE framework in use. We have deliberately selected cases that are applied, in the sense that they are relevant to industrial applications, as opposed to AI research.

### *Case 1: Predictive Maintenance*

Predictive Maintenance describes at its core the idea of optimizing the maintenance process by predicting and preventing machine failures proactively (Dalzochio et al. 2020; Lee et al. 2019; Hrnjica and Softic 2020). Found often in manufacturing environments but also in transportation or civil engineering, the goal of predictive maintenance is to minimize unplanned downtime, improve management of spare parts and repairs and enhance the overall equipment effectiveness (Dalzochio et al. 2020).

Different data sources can be used to bring the scenario to life. This can include machine condition data, historical maintenance records, real-time operational data from sensors or environmental factors like temperature or humidity. To use the thought frame introduced earlier, the goal is to infer and monitor the "health condition" of a machine by assessing different sensory inputs and historical data to anticipate and respond to potential problems (Lee et al. 2019). As the name suggests, predictive maintenance moves from reactive maintenance ("fix when it is broken") and periodic, scheduled maintenance towards using AI capabilities to predict machine failures (Hrnjica and Softic 2020). From a data perspective, the theoretical underlying concepts include time-series data from log-data or even sensory inputs. From an AI perspective, multiple approaches might be useful, from statistical analysis to more advanced models working on time-series data.

### *Case 2: Quality Control and Testing*

Often seen in manufacturing processes, quality control and testing with AI describes the idea of using AI-based solutions to automatically detect defects in an assembly line or inspect quality requirements at a scale. The goal is to increase the overall product quality and reduce defects through higher test coverage. Depending on the product and manufacturing process, various data sources can come into play, mapping to the steps a human inspector would test to ensure quality. Most commonly advanced computer vision algorithms are being used in visual inspection scenarios (Fahle, Prinz, and Kuhlenkötter 2020). Overall, the quality controls can also be embedded into a bigger picture using them as data inputs to predict the overall quality from design factors, manufacturing parameters, sensory data and quality checks (Tercan and Meisen 2022). Thus, the initial goal of quality control is assessing the quality or identifying defects but this can also be used to infer further actions.

### *RCAIE Framework*

Table 1 depicts the elements of the RCAIE framework for both cases introduced above. For each case, it lists how each of the constructs translates into discrete AI knowledge relevant for each of the roles: consumer, user, and creator and developer. The constructs were derived by mapping the relevant knowledge and skills for the use case and the respective role, in consultation with prior work. As presented here, the framework is an initial proposal that can be adapted by educators as needed and further validated by experts trough e.g. a Delphi Study.

*Table 1: The filled RCAIE framework with constructs for selected cases*

| # | Use Case | Roles | AI Literacy Constructs | | | | | |
|---|---|---|---|---|---|---|---|---|
| | | | Recognize | Know/ Understand | Use/ Apply | Evaluate | Create | Navigate Ethically |
| 1 | Predictive Maintenance | Consumer | Recognize that prediction relies on sensor data and AI systems | Understand the concept of predictive maintenance and how it supports the maintenance scheduling | Use maintenance recommendations to inform decision-making | Assess and compare predictions to own experiences and traditional approaches | - | Be aware of ethical implications and potential pitfalls in the use case |
| | | User | Recognize different types of sensor signals and predictions | Understand the concept of prediction and its connection to data | Use time series prediction on sensor data | Assess the impact and performance of different prediction algorithms | - | Consider ethical implications in the decision-making |
| | | Creator and Developer | Recognize different types of sensor signals and AI algorithms | Understand different types of sensor data and how they can be processed towards predictions using AI algorithms | Use different signal processing algorithms, build specific models using (physical) domain knowledge and deploy them | Test and validate different modelling approaches using multiple types of metrics | Develop new physical-based modelling approaches | Address and communicate ethical implications in the use case |
| 2 | Quality control and testing | Consumer | Recognize that automated quality control relies on sensor data and AI algorithms | Understand the difference between AI-based and traditional quality control | Use insights to make decisions about product rework or acceptance | Assess and compare quality control to own experiences and traditional approaches | - | Be aware of ethical implications and potential pitfalls in the use case |
| | | User | Recognize that AI capabilities useful for quality control | Understand different data sources, algorithm approaches and their limitations | Select and apply AI-based quality control systems for different manufacturing processes | Evaluate the performance in use | - | Ensure ethical and responsible deployment of tools |
| | | Creator and Developer | Recognize data potentials in manufacturing processes useful for quality control | Understand underlying data and model requirements for development and deployment | Use different AI capabilities based on the data requirements | Evaluate the effectiveness of models and overall solution with various metrics | Create new solutions and models for quality control | Advocate for responsible development and deployment |

## 4  DISCUSSION AND CONCLUSION

In this practice paper, we illustrate one approach for disciplinary engineering faculty to understand and integrate AI into their curriculum. As the description for each of the constructs in Table 1 shows, based on the area of application, different levels of knowledge need to be developed. From a basic and generic recognition of AI capabilities to an advanced understanding of models and training for a specific function, a range of options are available for learners. Their inclusion in curriculum depends on instructor expertise and training, as well as the level of course, and critically, students' prior knowledge.

### *Application of the Framework for Educators*

Given the increasingly broad and dynamic developments in AI, we argue that a roles-based approach to competency allows instructors to develop a mental model that provides them with an anchor such that they can start integrating AI aspects in a useful but simple manner. They can then keep building on constructs of competencies while keeping in mind the potential role of how an engineer would use AI and what additional knowledge they would need. Thus, educators can use the RCAIE framework as an approach to define and identify relevant competencies given identified use cases and future roles of their students. Moreover, a collection and synthesis of filled RCAIE frameworks in a domain can inform future development of curricula.

### *Limitations and Outlook*

Although we only use two case studies, it is easy to use the framework we have advanced to look at other disciplines or applications. The developed framework opens new avenues for research and has direct implications for the practice of engineering educators interested in integrating AI competencies in their studies.

We recognize that the framework has certain limitations. First, it is currently inductively developed from case studies only and, thus, not yet validated in use with educators. Second, the selection of case studies might have biased the development of the framework. In the future, these limitations can be overcome through a more thorough analysis of use cases, especially those that build on expert interviews to validate and deepen the insights into the domain-specific AI competencies. This approach can form the basis to identify domain-specific AI competencies and constructs that can be recognized across different engineering disciplines as well as other similarities and differences among the competencies. A collaborative approach with other educators so that they can share their identified competencies through an open repository will also help make this work stronger. Finally, competencies build the base for pedagogical approaches and this work can also be extended towards systemizing and analyzing the effectiveness of pedagogical approaches for different target groups or specific competencies.

**Acknowledgements**
This work is partly supported by U.S. NSF Awards# 2319137, 204863, USDA/NIFA Award#2021-67021-35329 and the German Federal Ministry of Education and Research under grant number 16DHBKI008. Any opinions, findings, and conclusions or recommendations expressed in this material are those of the authors and do not necessarily reflect the views of the funding agencies.